\documentclass[aps,pre,superscriptaddress,twocolumn,floatfix]{revtex4-1}
\usepackage{graphicx}
\usepackage{dcolumn}
\usepackage{bm}
\usepackage{hyperref}
\usepackage{amssymb}
\usepackage{amsmath}
\usepackage{array}
\usepackage[normalem]{ulem}

\usepackage{rotating} 

\usepackage{siunitx}

\usepackage{natbib}
\usepackage{epstopdf}

\usepackage{makecell}
\usepackage[table]{xcolor}
\usepackage{tabularx}
\usepackage{ragged2e}

\usepackage{lmodern}
\usepackage[latin1]{inputenc} 

\graphicspath{{figure/}} 

\definecolor{lightgray}{gray}{0.9}
\definecolor{darkgray}{gray}{0.60}
\newcolumntype{C}{c<{\kern\tabcolsep}<{\kern\tabcolsep}@{}}
\begin{document}
\clearpage

\newcommand{\be}{\begin{equation}}
\newcommand{\ee}{\end{equation}}
\newcommand{\del}{\partial}

\let\oldAA\AA
\renewcommand{\AA}{\text{\normalfont\oldAA}}

\newcommand{\LL}[1]{\textcolor{red}{{\bf LL:} #1}}
\newcommand{\MT}[1]{\textcolor{red}{{\bf MT:} #1}}
\newcommand{\Ch}[1]{\textcolor{cyan}{{\bf Ch:} #1}}
\newcommand{\LH}[1]{\textcolor{blue}{{\bf LH:} #1}}
\newcommand{\DF}[1]{\textcolor{blue}{{\bf DF:} #1}}
\newcommand{\JFJ}[1]{\textcolor{purple}{{\bf JFJ:} #1}}
\newcommand{\PBRO}[1]{\textcolor{cyan}{{\bf PBRO:} #1}}

\preprint{APS/123-QED}

\newcommand{\udem}{D\'{e}partement de Physique and Regroupement
 Qu\'{e}b\'{e}cois sur les Mat\'{e}riaux de Pointe, Universit\'{e} 
de Montr\'{e}al, C.P. 6128, Succursale Centre-Ville, Montr\'{e}al,
 Qu\'{e}bec, Canada H3C~3J7}
\newcommand{\nrc}{National Research Council of Canada, Ottawa, Ontario,
 Canada K1A 0R6}

\title{
Ion-ion dynamic structure factor, acoustic modes and equation of state of two-temperature warm dense aluminum
}

\author{L. Harbour}\email[Email address:\ ]{louis.harbour@umontreal.ca}\affiliation{\udem}
\author{G. D. F\"{o}rster}\affiliation{\udem}
\author{M. W. C. Dharma-wardana}\email[Email address:\ ]{chandre.dharma-wardana@nrc-cnrc.gc.ca}\affiliation{\nrc}
\author{Laurent J. Lewis}\email[Email address:\ ]{laurent.lewis@umontreal.ca}\affiliation{\udem}

\date{\today}

\begin{abstract} 
The ion-ion dynamical structure factor and the equation of state of warm dense
aluminum in a two-temperature quasi-equilibrium state, with the electron
temperature higher than the ion temperature, are investigated using
molecular-dynamics simulations based on ion-ion pair potentials constructed
from a neutral pseudoatom model. Such  pair
potentials based on density functional theory are parameter-free and 
depend directly on the electron
temperature and indirectly on the ion temperature, enabling
efficient computation of two-temperature properties.
Comparison  with \textit{ab initio} simulations and with other
 average-atom  calculations 
for equilibrium aluminum shows good agreement,  justifying
a study of quasi-equilibrium situations. 
Analyzing the van Hove
function, we find that ion-ion
correlations vanish in a time significantly smaller than the electron-ion
relaxation time so that dynamical properties have a physical meaning for the
quasi-equilibrium state. A significant increase in the speed of sound is
predicted from the modification of the dispersion relation of the ion acoustic
mode as the electron temperature is increased. The two-temperature equation of
state including the free energy, internal energy and pressure is also presented.
\end{abstract}

\maketitle

\section{Introduction}
The challenge of modeling warm dense matter (WDM) --- a system of strongly-coupled 
classical ions and partially degenerate electrons at high temperature
and high density --- is central to understanding many physical systems such as
the interior of giant planets \cite{Guillot99}, laser machining and ablation
\cite{Lewis03}, and inertial-confinement fusion \cite{Atzeni04}. In WDM, neither
the kinetic energy nor the potential energy can be treated as a perturbation.
Hence the usual theoretical techniques of classical plasma physics or
solid-state physics become inapplicable. In the laboratory, WDM can be created
through the interaction of high-energy short-pulse lasers with simple metals
such as aluminum \cite{Flet-Al-15, Ma-Al-13} and beryllium \cite{Lee-Be-09},
with densities $\rho$ several times the room density $\rho_0$ and temperatures
of the order of 1 eV. The importance of treating these systems as two-temperature
WDM systems rather than equilibrium systems has not always been
appreciated in analyzing the experimental results~\cite{cdwPlasmon, harbour16}. 

The ion-ion dynamic structure factor (DSF) $S(\mathbf{k},\omega)$ is a key
quantity for understanding the WDM regime. For instance, 
it contains information on the longitudinal waves propagating in the system.
The DSF can be measured by  neutron scattering and indirectly
via  X-ray Thomson scattering (XRTS) \cite{GlenzerRMP}. The Chihara decomposition \cite{Chihara} has been  applied 
to describe the XRTS signal by
partitioning
the total electron-electron DSF $S_{ee}(\mathbf{k},\omega)$ in the following form :
\be
S_{ee}(\mathbf{k},\omega) = S_{ee}^0(\mathbf{k},\omega) + N(\mathbf{k})S_{ii}(\mathbf{k},\omega) + S_{ee}^{fb}(\mathbf{k},\omega).
\ee
Here $S_{ee}^0$ is the free electron-electron DSF, $S_{ee}^{fb}$ is the contribution from
 transitions between bound and free electrons, $N(\mathbf{k}) = n_b(\mathbf{k}) + n_f(\mathbf{k})$
 is the total electron form factor split into a bound part $n_b$ and a free part $n_f$, and $S_{ii}$ is
 the ion-ion DSF. Using this decomposition, simulations combining standard  density functional
 theory (DFT) with molecular dynamics (MD) simulations have been used \cite{Flet-Al-15, Ma-Al-13, Ruter, Clerouin} to
 extract properties of WDM systems such as the free electron density per ion $n_e$ (i.e., $\bar{Z}$),
 the ion density $\rho$,
 the electron temperature $T_e$ and the ion temperature $T_i$. Since these simulations are computationally
 expensive, the use of a simpler approach such as the neutral pseudoatom (NPA) model is appropriate.
The NPA is  well adapted to extract the ion part of the XRTS signals \cite{harbour16} and will be
 the principal method used in this work, where we show that the NPA  also has the needed
accuracy. The XRTS spectrum can be computed in the NPA, and also
by other means without using the Chihara decomposition \cite{Vorberger15,Baczewski16} but its
 discussion is not needed for this work.

However, the separation between ion acoustic modes
$\Delta\omega = 2\hbar\omega_p$, with $\omega_p$ the ion-plasma frequency, is of
the order of 1 meV, significantly lower than the bandwidth of any X-ray probe
laser used experimentally at the moment. Thus, the ion-ion DSF is usually
approximated by its static form $S(k,\omega)= S(k)\delta(\omega)$ in describing
the XRTS signal. Nevertheless, the ion-ion DSF contains important information
about the ion transport properties linked to electron-ion equilibration, formation
of coupled modes, interaction with projectiles, etc.,  which makes
it a key quantity for fully understanding the WDM regime. With X-ray laser
sources being improved, the ion-ion DSF should become available from future
experiments, motivating its calculation for both equilibrium and quasi-equilibrium
situations. 

Furthermore, in the limit of small wavevectors, $k\rightarrow 0$, the ion-ion
DSF can be described in a hydrodynamic framework \cite{Hansen} (see below),
providing  important physical quantities such as the adiabatic
velocity of sound, the ion acoustic dispersion relation, the thermal diffusivity
and the sound attenuation coefficient. In addition, in the case of simple metals commonly
probed in most XRTS experiments, the laser interacts mainly with the
free-electron subsystem, creating a non-equilibrium system where the electron
temperature $T_e$ is higher than the initially cold ions at temperature $T_i$.
It has been shown that, when the shock wave resulting from the laser pulse has
propagated through the sample and reaches the probing location, the system might
still be in a two-temperature ($2T$) state \cite{harbour16, cdwPlasmon}.
 Since the ion-ion
interactions in simple metals are related to the screening of the free-electron
subsystem, the quasi-equilibrium properties of the total system with $T_e\ne T_i$ differ
significantly from the equilibrium ones. The hardening of the phonon spectra in
ultra-fast matter \cite{CPP-Harb,harbour17,Recoules06},  where $T_e$ is about 1 eV
 while $T_i$ remains at room temperature $T_r$, is an example of how the ion dynamics
 can be affected drastically in such conditions. Transport properties of Al in the
 two-temperature regime, such as self-diffusion and shear viscosity, are also
 significantly modified \cite{Hou17}.

The ion-ion DSFs for WDM have been calculated mainly using
DFT coupled to classical molecular dynamics (MD)
\cite{Ruter} simulations. Since  DSF calculations require a large
number of particles and long simulation times, DFT-MD calculations are
computationally very intensive. In addition, the finite-$T$ treatment of the electronic
subsystem in DFT requires the solution of the Kohn-Sham equations for
many electronic bands to take thermal excitations according to the Fermi-Dirac distribution
 into account. Orbital-free (OF)
 DFT simulations do not require
electronic wave functions, but require a Hohenberg-Kohn kinetic-energy functional
 as well as a finite-$T$ generalization thereof. Such procedures are less accurate than
the Kohn-Sham method, but make the simulations practical~\cite{White13}.
The full DFT-MD simulations when feasible can be used to benchmark  simpler methods
like the NPA or OF approaches which are easily applied over a wider range of
 temperatures  and densities.

In the present work, we compute the ion-ion DSF using classical MD simulations
based on pair potentials (PP) constructed from the  NPA model, which is a fully based on DFT.
 The NPA approach has  already been used to predict the DSF of strongly-coupled 
hydrogen plasmas \cite{NPA_DSF} and provides the
sound velocity, the thermal diffusivity, the specific-heat ratio, and the viscosity.
The NPA-PPs are free of {\em ad hoc} parameters and are
accurate  to within a few  meV as established by the prediction of  accurate experimental
phonon spectra for simple WDM solids~\cite{CPP-Harb, harbour17}. The NPA-PP-predictions for
 static structure factors (SSF) obtained using the
modified-hyper-netted-chain (MHNC) approximation are in agreement with DFT-MD simulations
 for WDM systems (typical  examples are Be and Al
\cite{harbour16}; for a review of the NPA, see Ref.~\onlinecite{CPP-cdw}).

Furthermore, in the quasi-equilibrium case, i.e.  when $T_e$ is  different
 from $T_i$, the extension
of the NPA approach to two-temperature ($2T$) situations has enabled the
construction of $2T$-PPs which reproduce \textit{ab initio} calculations of
quasi-equilibrium phonon spectra~\cite{harbour17}, quasi-equilibrium XRTS
signals~\cite{harbour16},  and frequency-dependent $2T$
 plasmon profiles~\cite{cdwPlasmon} and conductivities of ultra-fast matter \cite{Dharma17}.
The objective of the present study is to determine the DSF $S(k,\omega, T_e, T_i)$ in the $2T$ regime. In
addition, we evaluate the $2T$ equation of state (EOS). All calculations are
carried out for aluminum at the `room temperature'  density of $\rho_0$ = 2.7 g/cm$^3$ with
the ion temperature fixed at $T_i = 1 $ eV while the electron temperature $T_e$ is
varied between 1 and 10 eV.
\section{Methods}
\subsection{Neutral-pseudoatom model}
The NPA model \cite{Dagens1,Dagens2,NPA-PDW} is a rigorous all-electron
 DFT
average-atom approach where the ion distribution is also treated in DFT \cite{DWP1}.
 Given the mean free-electron density $n$ and
electron temperature $T_e$, it determines the total electron density around a
\textit{single Kohn-Sham ion} constructed from a nucleus of charge $Z_n$ embedded in
the plasma environment of mean density $\rho$. A classical Kohn-Sham equation for the
ions determines the one-body ion distribution $\rho(r)=\rho g_{ii}(r)$, where $g_{ii}(r)$
is the ion pair distribution function (PDF), abbreviated to $g(r)$. The
classical Kohn-Sham equation for the ions is identified as a type of hyper-netted chain (HNC)
integral equation bringing in ion-ion correlations beyond the mean-field approximation. 
The Kohn-Sham-Mermin solutions are obtained in the local-density approximation
using a finite-$T_e$ free-energy exchange-correlation (XC) functional $F_{xc}[n,T_e]$
\cite{PDWXC}. The available finite-$T$ XC-functionals, fitted to
quantum Monte-Carlo results or to the classical-map HNC results (used here),
yield numerically equivalent results in WDM applications~\cite{Hungary16}.  

In order to simulate the effect of the ion-density $\rho(r)$ on the electronic states, a 
uniform positive neutralizing background with a spherical cavity of radius $r_{\rm ws}$, with
the nucleus at the origin, is used. Here, 
$r_{\rm ws}=[3/(4\pi\rho)]^{1/3}$ is the Wigner-Seitz radius of the ion. This lowest-order model
for $\rho(r)=\rho g(r)$  is sufficient for calculating the Kohn-Sham energy levels of
``simple metal'' ions immersed in a warm dense electron fluid, as has been discussed
in a recent review \cite{CPP-cdw}. The adjustment of the ion distribution to the electron
distribution is accomplished by the optimization of a single parameter, viz.\ $r_{\rm ws}$,
subject to the finite-$T$ Friedel sum rule~\cite{DWP1}. Although, strictly speaking,  an electron-ion
exchange-correlation functional is also needed \cite{Furutani90}, it is neglected here.

An advantage of the NPA
model is that it directly provides single-ion properties such as the mean
ionization $\bar{Z}$ and the electron density around the nucleus
$n(r)=n_b(r)+n_f(r)$, with $n_b$ and $n_f$ the bound and free electron densities,
respectively. In simple metals, $n_b$ is found to be localized  
within a radius much smaller than $r_{\rm ws}$, such that $n_b(r\to r_{\rm ws}) = 0$, which
enables a clear definition of the mean ionization $\bar{Z} = n-n_b$. Note however, that
the free-electron distribution is {\it not} restricted to the WS-sphere, as is done in
many average-atom (AA) models, as reviewed  by, e.g., Murillo {\it et al.}~\cite{NPA-other5}. 

 The free electrons  occupy the whole space,  modeled  by  a large correlation sphere of
radius $R_c$  of about 10 WS radii, usually sufficient to include all particle
 correlations associated with the central nucleus.
 Unlike in AA models, the mean number of free electrons per ion, viz.\ $\bar{Z}$, is an unambiguously
 defined quantity subject to the
Friedel sum rule, and  experimentally measurable  using  XRTS~\cite{GlenzerRMP},
static conductivities, Langmuir probes, etc.
 The
interaction among ions of charge $\bar{Z}$ is screened by the free electron
subsystem which is assumed to respond linearly to the electron-ion
pseudopotential
\be
\label{pseudo.eq}
U_{ei}(k,T_e) = n_f(k)/\chi(k,T_e).
\ee
Since $n_f(k)$ is determined by the Kohn-Sham calculation which goes beyond the linear
response to $\bar{Z}/r$, the above pseudopotential actually includes all the non-linear DFT effects
within a linearized setting. The limits of validity of this procedure are discussed
in  Ref.~\cite{2Tpp}.

 With $n_f(k)$ at hand, $U_{ei}$ is constructed using the finite-$T$ interacting
 electron response function 
\be
\chi(k,T_e) = \frac{\chi_0(k,T_e)}{1-V(k)[1-G(k,T_e)]\chi_0(k,T_e)} 
\ee
with $\chi_0$ the finite-$T$ non-interacting Lindhard function, $V(k) = 4\pi/k^2$ the bare Coulomb interaction,
 and $G$ a finite-$T$ local-field correction \cite{CPP-Harb} which depends directly on $F_{xc}$. Finally,
 the screened ion-ion pair interaction is given by
\be
\label{pp.eq}
V_{ii}(k,T_e)= \bar{Z}^2V(k)+|U_{ei}(k,T_e)|^2\chi(k,T_e);
\ee
This pair potential is the NPA input to  the classical MD calculations.

It should be noted that the NPA uses a pair-potential for the ions and does not attempt to include
multi-ion potentials, as is customary in effective medium (EM) approaches that have been successfully
used for metals and semiconductors, especially at ambient temperature and compression. The EM  method
is at best a non-selfconsistent DFT approach~\cite{dePresito95} which includes two-body, three-body and
other multi-ion effects. It is often further extended by fitting
to empirical and calculational databases. However, recent attempts to use such  models for, e.g., WDM carbon, have not been very successful~\cite{Kraus13}.
The NPA approach  exploits the fact that  the grand potential $\Omega[n,\rho]$
is a functional of \textit{both} the one-electron distribution $n(r)$ and the one-ion distribution $\rho(r)$.
Hence a single-ion description (which allows a pair potential) is the only rigorously necessary information
 for a full DFT description of the system. In practice, pair potentials are sufficient if  linear-response
 pseudopotentials  could be constructed, as in Eq.~\ref{pseudo.eq}.
 However, this approach now needs, not only an XC-functional for the electrons,  but
also a correlation functional for the classical ions. These are constructed via classical integral equations, or
automatically via MD simulations. Detailed discussions of these issues and the  NPA method may be found
 in Refs.~\cite{Ilciacco93,CPP-cdw}.
In this context, we remark that standard implementations of DFT-MD in codes like
ABINIT and  VASP \cite{Abinit, VASP} use only the one-electron
 density-functional
property, and {\em not} the one-ion density functional, as it chooses to implement a
 full $N$-ion Kohn-Sham simulation
with $N$ typically of the order of 100 or more. 

Furthermore, the multi-center nature of the simulations implies a
highly non-uniform electron density requiring sophisticated gradient-corrected XC-functionals. In contrast, the
NPA uses  a relatively smooth single-center electron distribution for which the local-density approximation (LDA) is
found to work very well, even for sensitive properties like the electrical conductivity~\cite{Dharma17}
and plasmon spectral line shapes~\cite{cdwPlasmon}. The LDA form of the finite-$T$ XC-functional of Perrot and
Dharma-wardana~\cite{PDWXC} is used  in this study.

\subsection{Dynamic Structure Factor}
\label{Method.DSF}
The ion-ion spatial and temporal correlations are determined from the van Hove function 
\begin{align}
G(\mathbf{r},t)& = \frac{\langle \rho(\mathbf{r},t)\rho(\mathbf{0},0)\rangle}{\rho} \\
&\nonumber = \frac{1}{N} \left\langle \sum_{i=1}^N \sum_{j=1}^N  \delta[\mathbf{r}-\mathbf{r}_j(0)-\mathbf{r}_i(t)]\right\rangle
\end{align}
with $\langle\cdot\cdot\cdot \rangle$ the ensemble and time 
 average
(over many different time origins)
 calculated from classical MD simulations of  a $N$-particle system, $\rho$ the mean ion density, and $\mathbf{r}_i(t)$
  the position of the $i$-th ion at time $t$.  The ion-ion DSF  
\be
S(\mathbf{k},\omega)=\frac{1}{2\pi}\int_{-\infty}^\infty F(\mathbf{k},t)\ e^{i\omega t}dt
\ee
is the time Fourier transform of the intermediate scattering function $F(\mathbf{k},t)$, which is
 itself the spatial Fourier transform of the Van Hove function 
\be
F(\mathbf{k},t) = \int G(\mathbf{r},t)e^{-i\mathbf{k}\cdot\mathbf{r}} d\mathbf{r}.
\ee
While $G(\mathbf{r},t)$ contains much information relevant to $2T$ situations, 
 $F(\mathbf{k},t)$ is also directly accessible in MD simulations via the relation 
\be
F(\mathbf{k},t) =\frac{1}{N} \langle\rho_\mathbf{k}(t)\rho_{-\mathbf{k}}(0)\ \rangle
\ee
where
\be
\rho_\mathbf{k}(t)=
 \sum_i^N e^{i\mathbf{k}\cdot \mathbf{r}_i(t)},
\ee
thus avoiding the calculation of the spatial Fourier transform which can add spurious high-frequency
 oscillations to $F(\mathbf{k},t)$ due to the finite size of
the MD simulation cell. Under  WDM conditions, the averaged system properties are
 those of an  isotropic fluid; thus important structural quantities are spherically-symmetric in real space,
 $|\mathbf{r}| = r$, and in reciprocal space, $|\mathbf{k}|=k$. 

In the hydrodynamic limit, $k\to0$, the DSF takes the so-called `three-peak' form
\begin{align}
\label{hydroskw}
&S(k,\omega)= \frac{S(k)}{2\pi}\left[\left(\frac{\gamma-1}{\gamma}\right)\frac{2D_Tk^2}{\omega^2+(D_Tk^2)^2}\right. \\
&\nonumber +\left.\frac{1}{\gamma}\left(\frac{\Gamma k^2}{(\omega-c_sk)^2+(\Gamma k^2)^2} +\frac{\Gamma k^2}{(\omega+c_sk)^2+(\Gamma k^2)^2}\right) \right]
\end{align}
with $D_T$ the thermal diffusivity, $\Gamma$ the sound attenuation coefficient, $\gamma=c_P/c_V$ the ratio of the constant
 pressure to the constant volume specific heats, $c_p$ and $c_V$, and $c_s$ adiabatic speed of sound. The second and third
 terms of Eq.~\ref{hydroskw} are the Brillouin peaks whose positions provide the acoustic dispersion relation $\omega_s(k)$,
 which is linear at small $k$, $\omega_s(k\to 0) = c_sk$, and is measurable experimentally. In addition, it is also
 possible to compute $c_s$ from the SSF $S(k)$ using the compressibility $\kappa$ sum rule $S(0) = \rho\kappa T_i$ which leads to  $c_s = \sqrt{T_i/S(0)} $. Once the PP is constructed,
 the SSF can be easily calculated via the MHNC procedure, independent of MD simulations.
\subsection{Equation of State}
The total free energy per atom in the NPA model is given by
\be
\label{FreeEnergy}
F = F_e^0(T_e) + F_\text{emb}(T_e) + F_{ii}(T_i,T_e) + F_i^0(T_i),
\ee
with contributions $F_e^0$ from the interacting homogeneous electron gas, $F_ \text{emb}$ from the embedding of the
 pseudoatom into the uniform system, $F_{ii}$ from the interacting ion-ion system, and $F_i^0$ from the ideal ion gas.
  A more detailed description of each term of the NPA free energy is given in Refs.\ \cite{NPA-PDW, eos95}.
 For the equilibrium system,
 the pressure is obtained via the density derivative of the free energy while the internal energy is obtained by
 taking the  temperature derivative:
\be
P = n^2 \frac{\del F}{\del n}, \qquad
E = \frac{\del (\beta F)}{\del \beta}
\ee
with $\beta = 1/T$. For quasi-equilibrium systems, the internal energy must be computed taking into account the
 temperature derivative of each contribution in Eq.\ref{FreeEnergy}. Note that the term $F_{ii}$ depends on both
 $T_i$ and $T_e$. Thus, the total derivative of the  $2T$ internal energy reads
\be
E(T_e,T_i)\equiv \frac{\del(\mathbf{\beta}F)}{\del\mathbf{\beta}} = F + \left.\frac{\del F}{\del \beta_e}\right|_{T_i}
 + \left.\frac{\del F}{\del \beta_i}\right|_{T_e}
\ee
which recovers the correct equilibrium internal energy when $T_i = T_e$.
\section{Results}
All DSF have been calculated from MD simulations using the NPA pair potentials.
 The initial configuration was a face-centered cubic crystal
 containing  5324 particles arranged in a cubic simulation cell. This corresponds approximately to a linear dimension
of 17 to 18 Wigner-Seitz radii, i.e., significantly larger than typical ion-ion correlations seen in the
 ion-ion pair distribution of aluminum even at its melting point.
 Simulations were carried out over 0.5 ns with a timestep of 0.5 fs. 
The first  50,000 steps have not
 been used as they pertain to the initial equilibration period. From the remaining simulation, configurations 
have been extracted every 1 fs for the
 calculation of $G(r,t)$ and $F(k,t)$ which have been calculated up to
 3000 fs. The
ion  temperature was kept constant throughout the simulation using a Nos\'{e}-Hoover thermostat.
 The electron temperature no
longer appears in the dynamics, so no electron thermostat is needed; $T_e$ only intervenes in the construction
 of the NPA pair potential, which is the essential `quantum input' to the classical MD simulation. 

In the range of $T_e$ studied in this work, i.e. from 1 to 10 eV, the mean ionization calculated
 from the NPA model remained essentially unchanged from the room temperature value of $\bar{Z} = 3.0000$
to 3.0163  for the normal density of 2.7 g/cm$^3$. Given a Fermi energy (i.e, approximately the chemical potential) 
of  $\sim$ 12 eV, no significant change in $\bar{Z}$ is in fact expected. There is even less of a
change at the higher density of 5.2 g/cm$^3$ used by R\"{u}ter {\em et al} \cite{Ruter}, as the Fermi energy
 is correspondingly higher. The value of  $\bar{Z}$ for Al begins to
 increase only from about 20 eV,
and the consistency of the NPA evaluated $\bar Z$ even at higher temperatures is shown from its successful
 prediction of electrical conductivities  of aluminum under a variety of WDM conditions \cite{Dharma17,Benage00}.

\subsection{Static properties}
We first review the results for several key static properties, viz.\  pair-potentials, PDFs, and structure factors.

\subsubsection{Pair-potentials}
\begin{figure}[h]
\includegraphics[width=\columnwidth]{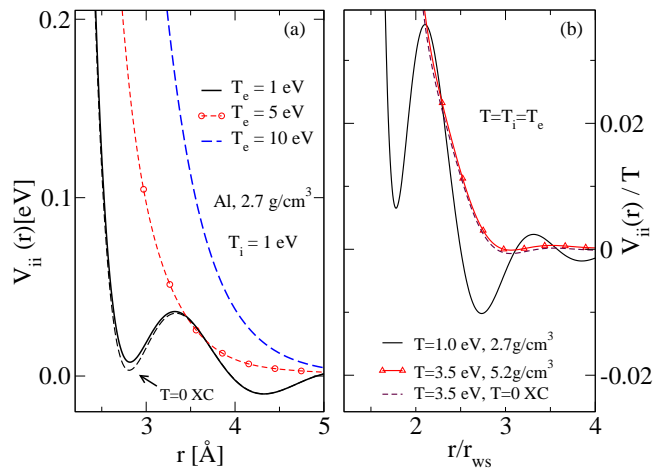}
\caption{(color online) (a) Ion-ion pair potentials constructed from the NPA model at electron 
temperatures of $T_e =$ 1, 5,
 and 10 eV, while the ion temperature is held at $T_i$ = 1 eV. (b) The
$T_i=T_e=1$ eV potential for the `room temperature' Al  density 2.7 g/cm$^3$,
 as well the pair potential at 5.2 g/cm$^3$ and T=3.5 eV relevant  to the work
  of R\"{u}ter R\"{u}tter and Redmer \cite{Ruter}. Note that in panel (b)  we have plotted the potentials in terms of physically
relevant variables $r/r_{\rm ws}$ and $V_{ii}(r)/T$, where the nominal WS radii $r_{\rm ws}$ are
 2.991 a.u.  and 2.404 a.u, respectively.
}
\label{fig:2T.Vr}
\end{figure}

The  easily computed NPA ion-ion pair-potentials described by Eq.~\ref{pp.eq} are the starting point of our study
of the aluminum DSF, using classical molecular dynamics with the  NPA-PPs as the input. Hence, in Fig.~\ref{fig:2T.Vr}
we show typical  Al-Al pair potentials that are relevant to our study. These pair potentials are the simplest
that can be constructed from the NPA density, while the NPA calculation provides enough data to construct more
complex non-local pseudopotentials, or potentials designed to recover phase shifts, etc. However, such elaborations
 need to be invoked only if such potentials are really required. We have found that this elementary approach works
 well for simple metallic fluids in regimes of  compressions of 0.5 to about 2.5, and from low temperatures (e.g.,
melting point) to  higher temperatures (where the model works better).
In the present study (aluminum at 2.7 g/cm$^3$, and 5.2  g/cm$^3$, at $T$=1 eV and 5.4 eV, the model is eminently
 applicable, as we show by comparisons with more microscopic simulations for the PDFs and other properties given below. 
Panels (a), (b) show the crucial role played by the Friedel oscillations in the potentials. These are evident
in the potential at $T_e=1$ eV and more weakly in the 3.5 eV potential. At high $T_e$, they are damped out
and the potentials become more Yukawa-like.  The NPA model faithfully reproduces these
potentials to good accuracy, whereas many commonly-used  average-atom models do not. The location of these
oscillations, as well as packing effects in the fluid, are controlled by $r_{\rm ws}$.  Hence the plot
using $r/r_{\rm ws}$ as the $x$-coordinate brings potentials at different densities to a comparable footing.
We also show the pair-potential at 5.2 g/cm$^3$ and $T$=3.5~eV calculated using the $T$=0 XC-functional
that is customarily implemented in DFT-MD simulations, showing a small and probably negligible difference. However,
it should be always remembered that standard DFT-MD simulations can be used to benchmark other calculations only
when the $T=0$ XC approximation holds.
  
\subsubsection{Pair Distribution Function} 
The NPA-PPs are  known to closely reproduce the ion-ion PDF $g(r)$ and the corresponding static structure
 factor $S(k)$ for most systems  studied so far, for compressions of 0.5 to about 2.5. Some examples are : \\ 
(i) Al (a) at normal density $\rho = 2.7 $g/cm$^3$ and at the melting point, and (b) at an expanded density $\rho = 2.0$ g/cm$^3$ with $T = $ 1,000K and 5,000K   \cite{Dharma06},\\
(ii) Li at $T = 2000$ K and $\rho = 0.85$ g/cm$^3$\cite{harbour17}, \\
(iii) Be at densities of $\rho =1.85$ g/cm$^3$ and $\rho = 5.53$ g/cm$^3$ for various two-temperature situations \cite{harbour16}, and \\
(iv)  C, Si and Ge in the WDM state \cite{Lqd-C2016,DWP-Carb90}. Here, because of the high
 electron density ($\bar{Z}=4$) the NPA model works even at 12 g/cm$^3$, i.e, close to six times the graphite density.

 While liquid metal PDFs can be  obtained  from MD simulations using
 multi-center potentials such as those available from EM theory~\cite{dePresito95},  
embedded-atom model (EAM) approaches \cite{EAM-Al},
 or  bond-order potentials \cite{ghiring05}, they have not been applied in an intensive way to the WDM regime.
 The effect of  $T_e$  on the ion-ion interaction is  not
 included except in limited  cases \cite{Khakshouri08}. Kraus {\em et al}. \cite{Kraus13} examined the use of
 multi-ion  bond-order potentials for WDM carbon but found them to be unsuitable and extremely difficult
 to formulate for finite-$T$ usage. In contrast, the NPA-PPs are simple to compute and are at finite-$T$ from
 the outset. Here we  show that the PDFs obtained from them agree closely with those from DFT-MD for the
 systems studied here. 
\begin{figure}[h]
\includegraphics[width=1\columnwidth]{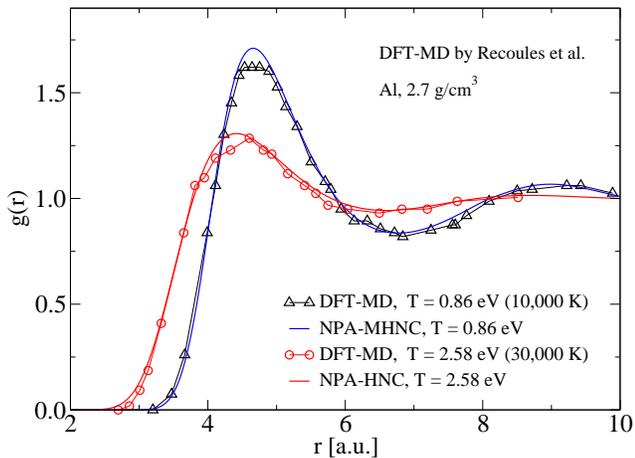}
\caption{(color online) A comparison of the $g(r)$ from the NPA-MD  and from DFT-MD simulations 
(Recoules {\em et al}. \cite{Recoules-gr15}
 for  Al at the normal density $\rho_0$ and at 10,000K and 30,000K. No bridge terms are needed at the higher $T$ case,
where HNC and MHNC become equivalent).}
\label{Recou-gr.fig}
\end{figure}

\begin{figure}[t]
\includegraphics[width=\columnwidth]{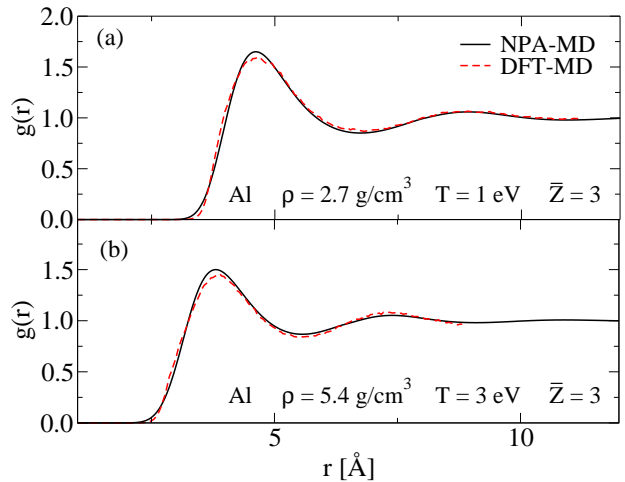}
\caption{(color online) Comparison of the $g(r)$ from the NPA-MD  and from DFT-MD simulations
 for two Al WDM states:  (a) $\rho$ at the room temperature density of 2.7 g/cm$^3$, $T = 1$~eV, and
 (b) $\rho = 5.4$ g/cm$^3$, $T=3$~eV.}
\label{fig:gr_NPA_DFTMD}
\end{figure}

In  Fig.\ref{Recou-gr.fig} NPA PDFs for aluminum at the `normal' density $\rho_0$  are compared with
DFT-MD simulations from Recoules \textit{et al} at 10,000 K and 30,000 K \cite{Recoules-gr15}. The agreement is relatively good.
 The slight disagreement ($\sim$ 4\%) noted near the main peak is a common feature in this type of comparison,
 arising from statistical noise in MD simulations with, say $N \sim$ 100 atoms.  Here one may
expect fluctuations of $\sim 1/\surd{N}$. In reality, the need to take an ensemble average of every quantity
in DFT-MD simulations adds to the labour and cost. In  Fig.\ref{fig:gr_NPA_DFTMD}, we compare the PDF for Al obtained from the NPA pair potentials with that from Kohn-Sham DFT-MD simulations for two cases. The first case (panel (a) in Fig.\ref{fig:gr_NPA_DFTMD}) is for the room temperature density $\rho = 2.7$ g/cm$^3$ at a temperature $T=1$ eV, which gives one of the equilibrium WDMs used in this study. The second case (panel (b) in Fig.\ref{fig:gr_NPA_DFTMD}) is for the compressed density $\rho = 5.4$ g/cm$^3$ at  a temperature 
 $T = 3$ eV which is close to the conditions used  by  R\"{u}ter and Redmer \cite{Ruter} in their DFT-MD calculation of the aluminum DSF. The latter is used in section \ref{Sec.DSF} to compare with our NPA-MD DSF. Our DFT-MD simulations were done with the ABINIT package  using a cell of 108 atoms with a norm-conserving pseudopotential and the $T=0$ Perdew-Burke-Ernzerhof exchange and correlation functional within the generalized gradient approximation. In this case, the position of the first maxima in $g(r)$ are within  1\%  of each other for the first case (a) and within 2\% at for the second case (b). The height of the first peak differs by about 3\% in both cases showing the good aggreement between DFT-MD and NPA-MD simulations. The use of pair potentials to perform classical MD simulations requires a considerably shorter amount of time illustrating the advantage of employing the NPA model. 
 
\subsubsection{Static Structure Factor} 
 
\begin{figure}[h]
\includegraphics[width=\columnwidth]{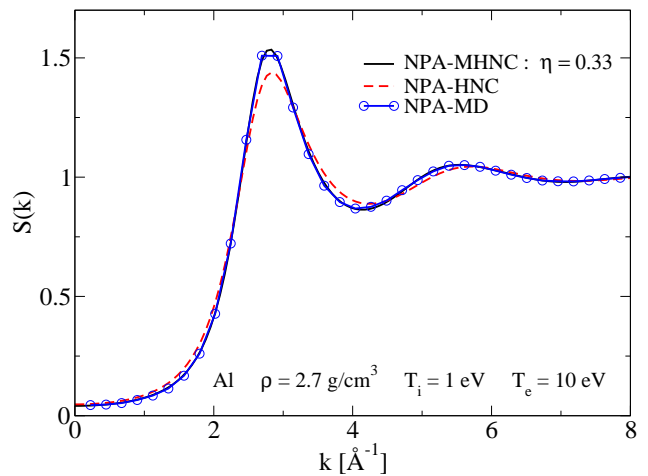}
\caption{(color online) $S(k)$ computed from MHNC (continuous line), HNC (red dashed line) and MD (blue circles and blue line)
 simulations at $T_i$=1 eV and $T_e$=10 eV.}
\label{fig:gr_hnc-MHNC}
\end{figure}

 As indicated in Sec.\ref{Method.DSF}, $c_s$ can also be calculated
 from the SSF using the compressibility sum rule.  In Fig.~\ref{fig:gr_hnc-MHNC}, we  compare the $S(k)$ computed
 from HNC, MHNC and MD simulations all using the same pair potential. We note that the MHNC SSF and  the  MD SSF
agree very well,  while the HNC predicts a slightly lower maximum and a slightly different $k=0$ limit. This
suggests that the differences may be due to the use of a  hard-sphere model within the Lado-Foiles-Ashcroft criterion
for modeling the bridge function \cite{LFA}. In principle, more accurate bridge functions can be extracted
from MD simulations.
\subsection{Dynamical properties}\label{Sec.DSF}
\subsubsection{Dynamical Structure Factor: Equilibrium system}
\begin{figure}[h]
\includegraphics[width=\columnwidth]{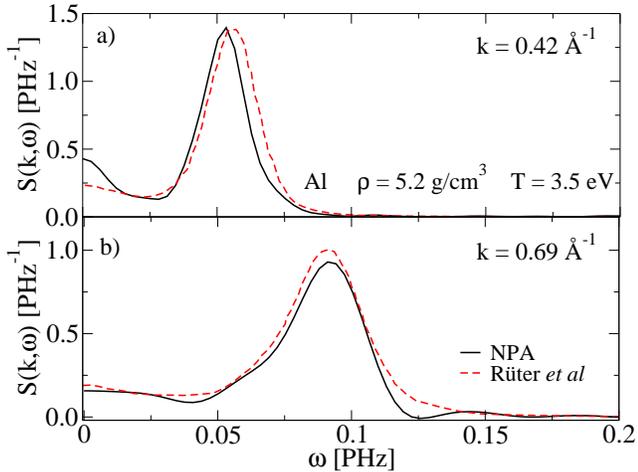}
\caption{A comparison of the dynamic structure factors obtained from NPA-MD (this work) and DFT-MD \cite{Ruter}
 at two different wavevectors. In NPA-MD, the pair potential (Eq. \ref{pp.eq}) is the input to simulation.}
\label{fig:DSF_Ruter}
\end{figure}

In this section, the DSF for  equilibrium Al,  $T_i = T_e$, obtained from the NPA-PP, is compared
 with other DSF calculations. First we consider the results of R\"{u}ter and Redmer \cite{Ruter} who
 used  Kohn-Sham DFT-MD to study Al at a density of $\rho = 5.2$ g/cm$^3$ (compression $\sim 2$) and
  $T = 3.5$ eV, i.e., $T/E_f\simeq 0.19$.
 In Fig.\ref{fig:DSF_Ruter}, the NPA-MD   DSF is compared with the DFT-MD DSF  for  $k = 0.42\  \AA^{-1}$
 and $k = 0.69\ \AA^{-1}$.  The position and profile of the Brillouin peak obtained from the NPA-MD agree
 closely with results from DFT-MD. Furthermore,
 the speed of sound obtained by R\"{u}ter and Redmer, $c_s = 10.38$ km/s, and  the NPA value of $c_s = 10.62$ km/s
are within 2.3\% of each other. In this case, the NPA calculation satisfies the $f$-sum rule to within 96$\%$ over the
 range of $k$ studied.

\begin{figure}[h]
\includegraphics[width=\columnwidth]{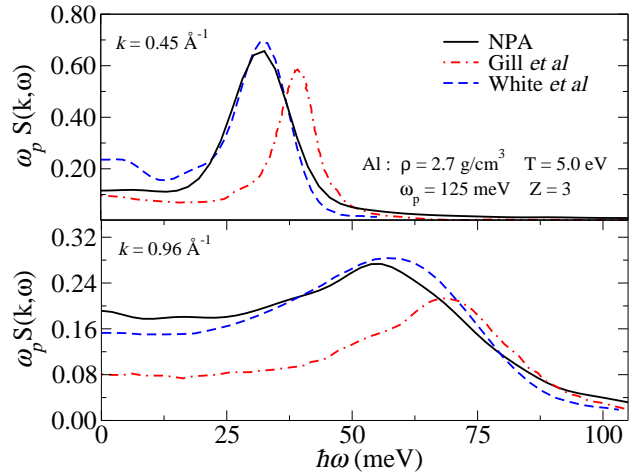}
\caption{(color online) The equilibrium DSF of Al at  
density $\rho = \rho_0$  and $T$ = 5 eV,  for
 wavevector $k=0.45\ \AA^{-1}$ (upper panel) and
 0.96 $\AA^{-1}$ (lower panel): NPA (black continuous line), PA-SS~\cite{gill15} (red dot-dashed line)
 and OF-DFT-MD~\cite{White13} (blue dashed line).
}
\label{fig:Eq.DSF.T5}
\end{figure}

Since a full Kohn-Sham DFT-MD calculation as  given by R\"{u}ter {\em et al.} \cite{Ruter}
for the the DSF is extremely costly, simpler approaches
 based on average-atom models as well as orbital-free methods have been used  to compute the ion-ion DSF.
 Here  we  compare the results from the NPA-MD with corresponding results from 
 the pseudoatom model of Starrett and Saumon \cite{NPA-other1}(PA-SS), and with OF-DFT-MD simulations, 
 for   Al at the  density  $\rho_0$  and $T$ = 5 eV.
 A comparison of our NPA-MD
 calculations with
 the OF-DFT-MD simulations of White \textit{et al}. \cite{White13} and those of Gill \textit{et al}. \cite{gill15},
 using the  PA-SS and MD,  is presented in Fig.~\ref{fig:Eq.DSF.T5} for wavevectors $k=0.45\ \AA^{-1}$ 
 and 0.96 $\AA^{-1}$.  White \textit{ et al.} used  108 ions in a cubic supercell  in an
 OF-DFT approach. Gill \textit{et al.} 
presented an OF model calculation with a classical simulation with 10,000 ions, and also a
 Kohn-Sham (KS) approach within their PA-SS model.
Since our NPA model uses the KS procedure, only the KS-PA-SS results
 are compared in Fig.~\ref{fig:Eq.DSF.T5}.

The positions of the Brillouin peak
for $k$ = 0.45 $\AA^{-1}$ coincide for OF-DFT-MD and NPA-MD, and the peak heights
differ by $\sim$4\%. The adiabatic speed of sound $c_s =\omega_s / k$ was obtained by a linear
 fit to the dispersion  relation $\omega_s(k)$ for small value of $k$. The OF-DFT-MD predicts an adiabatic
 speed of sound  of 10.4 km/s, very close to the NPA-MD value of 10.2 km/s
 whereas the PA-SS-MD predicts a higher value of 12.7 km/s. Once again, we ensured that  the NPA calculation satisfies the $f$-sum rule to within 97$\%$ over the
 range of $k$ studied. The good  agreement between
 the equilibrium DSF  calculated via the NPA-MD and OF-DFT-MD  mutually confirm the extent of validity
 of these methods and of  the  NPA-MD approach. We already noted the
 good agreement with the
fully microscopic calculations of R\"{u}ter {\em et al} \cite{Ruter}. All these encourage us to apply
 $2T$-NPA-PP to investigate dynamical properties of  quasi-equilibrium
 systems where~$T_e\ne T_i$.
 
 \begin{figure}[h]
\includegraphics[width=\columnwidth]{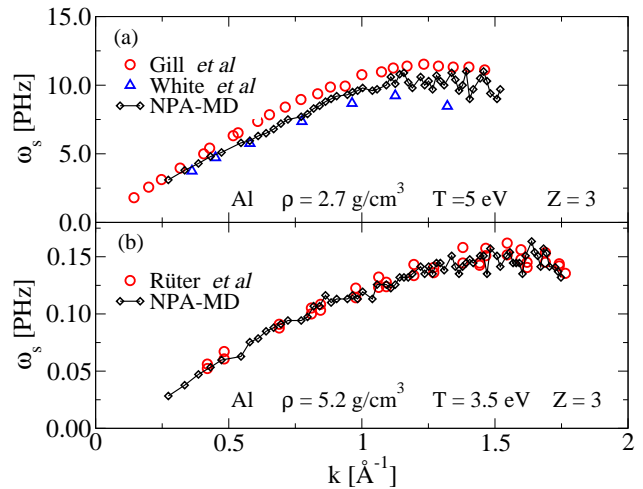}
\caption{(color online) A comparison  of the equilibrium
 acoustic dispersion relation of Al for (a)  density 
 $\rho=\rho_0$ and $T$ = 5 eV
as in the OF-DFT-MD calculations of White {\em et al} \cite{White13} and as in the PA-SS calculations of Gill {\em et al} \cite{gill15}, and   
(b)  $\rho = 5.2$~g/cm$^3$ and $T$ = 3.5 eV as in R\"{u}ter {\em et al} \cite{Ruter}.}
\label{fig:Eq.Disp}
\end{figure}
\subsubsection{Dynamical Structure Factor: Quasi-equilibrium system}
In order to study the ion dynamics in the quasi-equilibrium system with $T_e > T_i = 1$ eV,  we employ
 the ion-ion pair potential  $V_{ii}(r,T_e)$ constructed from the NPA calculation which explicitly depends
directly on $T_e$. The dependence on $T_i$ comes in via
 the ion density and the ionization state $\bar{Z}$ of the ions, and hence is implicitly
included in the  NPA calculation. In Fig.~\ref{fig:2T.Vr}(a) we present
 the potentials for the cases $T_e =$ 1, 5, and 10 eV. At $T_e = 1$ eV, the potential
 exhibits Friedel oscillations
with several minima, whereas it becomes purely repulsive at higher temperatures since the Fermi energy at
2.7 g/cm$^3$ is 11.65 eV.

\begin{figure}[h]
\includegraphics[width=\columnwidth]{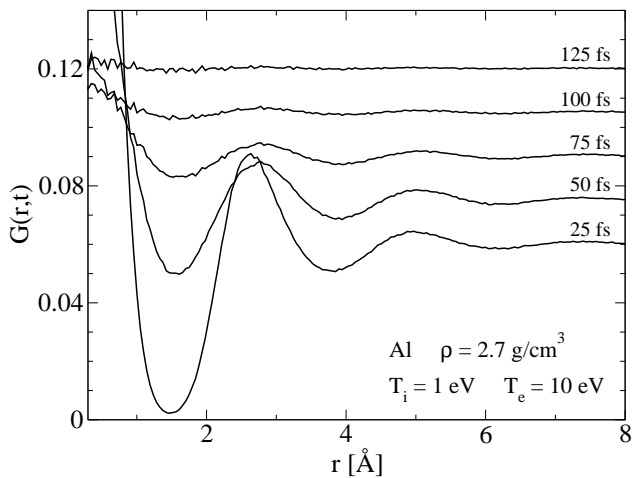}
\caption{The Van Hove function for different times for the  case $T_i= 1$ eV and $T_e = 10$ eV. 
For clarity, each curve is shifted vertically by 0.015 from the previous one while the curve for $t=25$ fs
is unshifted.
}
\label{fig:2T.Van}
\end{figure}

To ensure that the two-temperature DSF of the quasi-equilibrium system is physically relevant, 
we must verify that all spatial correlations vanish in a time $\tau_c$ smaller than the ion-electron
 relaxation time $\tau_{ei}$,
 which is of the order of hundreds of picoseconds \cite{elr2001}. The Van Hove function has been calculated
 for the specific  case $T_i=1$~eV and $T_e=10$~eV and its time evolution  is presented in
  Fig.~\ref{fig:2T.Van}.
  We find that at
 $\tau_c = 125$ ps, all spatial correlations have vanished, such that $\tau_c < \tau_{ei}$,
  implying that dynamical
 properties can be meaningfully calculated for the quasi-equilibrium system.

\begin{figure}[h]
\includegraphics[width=\columnwidth]{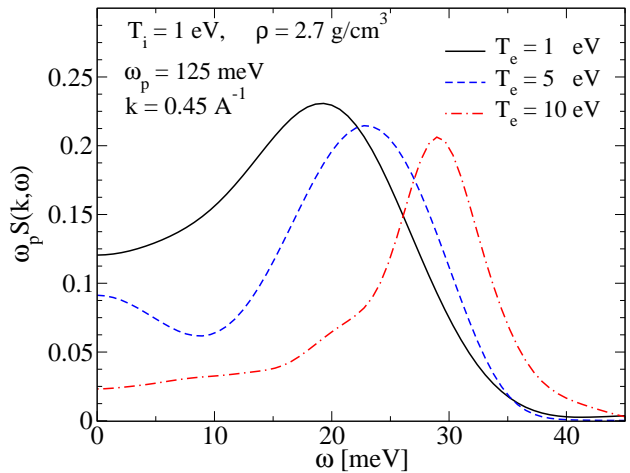}
\caption{(color online) The quasi-equilibrium DSF of Al at  density $\rho_0$, $T_i$ = 1 eV
 and $T_e = 1, 5$, $T$= 10 eV, for wavevector $k=0.45\ \AA^{-1}$.}
\label{fig:2T.DSF}
\end{figure}

The $2T$-DSF at $\rho=\rho_0$,
 computed with the NPA-based PP, is presented in Fig.~\ref{fig:2T.DSF} for
 $T_i = 1$ eV  and $T_e= 1, 5$ and 10 eV and wavevector $k = 0.45$ $\AA^{-1}$.
 The position of the
 Brillouin peak shifts to higher
 $\omega$ as $T_e$ increases while the value at $\omega = 0$ is drastically lowered.
 Furthermore, the shape of
 the peak  is narrower for higher $T_e$. 

\begin{figure}[h]
\includegraphics[width=\columnwidth]{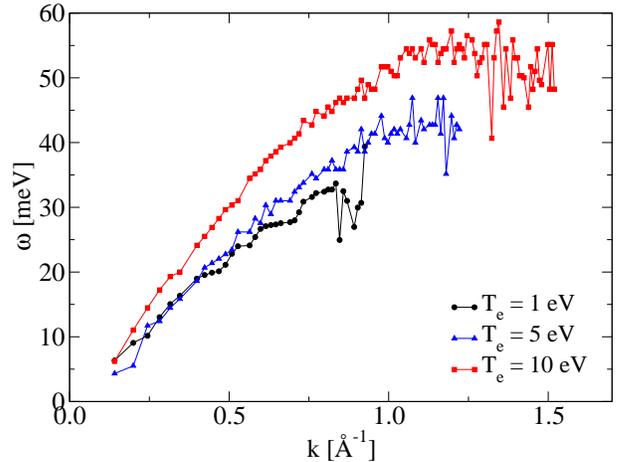}
\caption{(color online) The two-temperature dispersion relation $\omega_s(k)$ calculated from
 the position of the Brillouin peak for $T_i = 1$ eV and $T_e =$ 1, 5,
 and 10 eV.}
\label{fig:2T.dispersion}
\end{figure}

The  dispersion relation $\omega_s(k)$  for $T_e = 1, 5$ and
 $10$ eV can be deduced from the position of the Brillouin peak. It is displayed
 in Fig.~\ref{fig:2T.dispersion}.

\begin{figure}[h]
\includegraphics[width=\columnwidth]{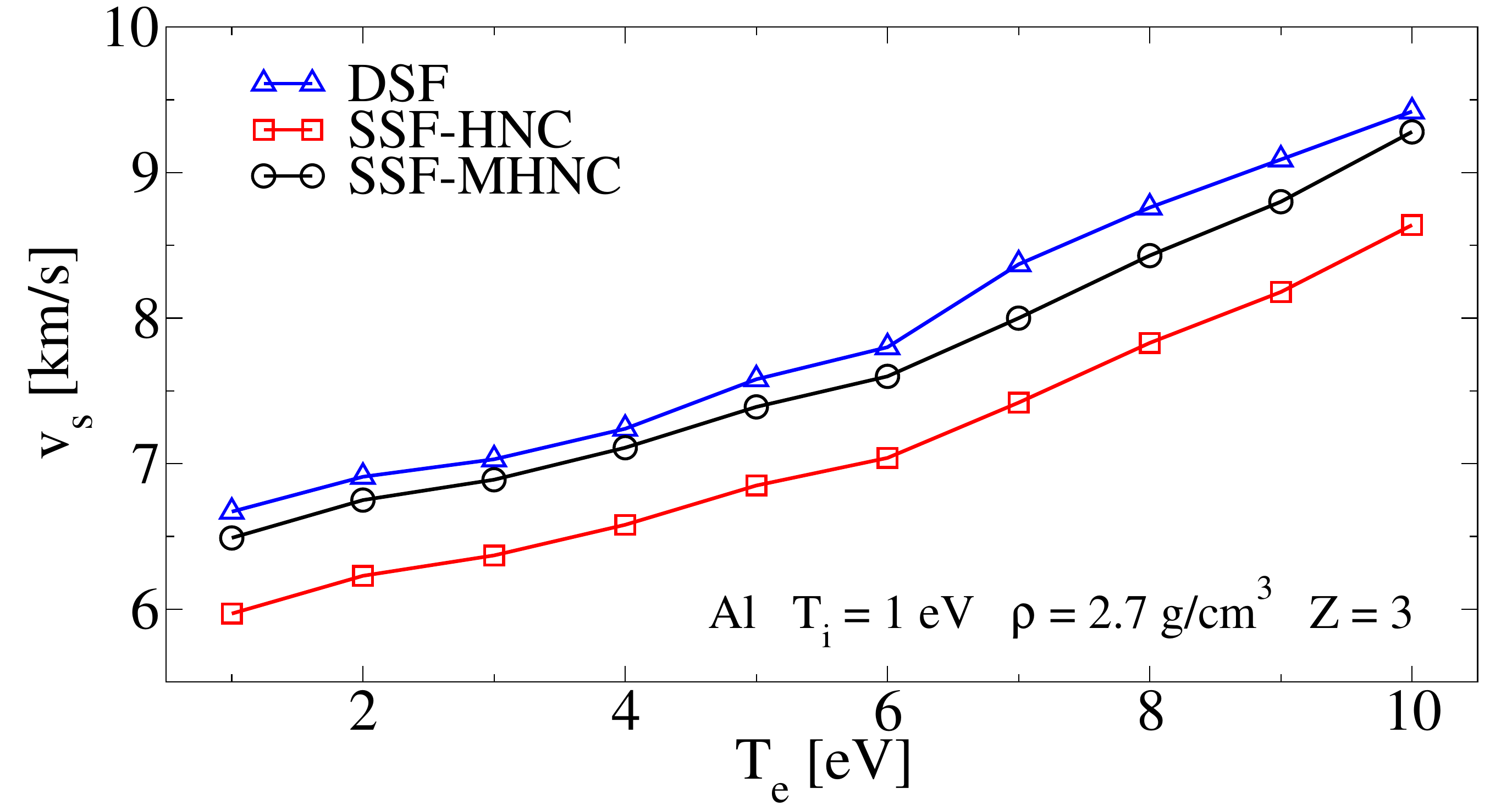}
\caption{(color online) Comparison of the speed of sound in the $2T$-system calculated from the SSF
 and from the DSF with $T_i =  1$ eV.}
\label{fig:2T.cs}
\end{figure}

The dispersion relation begins to be noisy and unphysical at different values of wavevector $k$ as 
  $T_e$ is increased. Thus the  position  of the Brillouin peak could be confidently determined
 only up to $k = $ 0.8, 1.2, and 1.5 $\AA^{-1}$  for $T_e =$ 1, 5, and 10 eV,
 respectively. Establishing that collective excitations still exist for higher values of $k$
 becomes more difficult as the Brillouin peak merges back with the Rayleigh peak at $k=0$. This makes it
 hard to  evaluate the full width at half maximum of the Brillouin peak,  ideally needed
 to establish  the survival of longitudinal modes  at higher $k$ and  higher $T_e$. Instead, we decided
 to include the position of the peak as long as its height is at least 20$\%$ higher than the value at $k=0$.
 Using the same procedure for each combination of $T_i$ and $T_e$ enables us to treat them in a comparable manner.
  Longer MD simulations would yield better results for $\omega_s(k)$; however the current results are sufficiently
 precise to conclude that there exist more ion longitudinal modes at $T_e =$ 10~eV than at lower $T_e$;
 this may be due to the lower compressibility of the electron subsystem as well as the ion subsystem
 with a   more repulsive pair potential, as shown in Fig.~\ref{fig:2T.Vr}, the ion temperature being identical.
 These dispersion relations will be used to determine the speed of sound $c_s$ as a function of $T_e$.
  Predictions of the speed of sound computed from the DSF, the MHNC-SSF and the HNC-SSF are
 compared in Fig.~\ref{fig:2T.cs}.

The speed of sound calculated from the DSF is slightly and systematically higher than the MHNC-SSF value
 through the entire range
 of $T_e$, with a maximum difference of 4.6$\%$ occurring at $T_e = 7$ eV;  both methods predict a 43$\%$ increase
 from  $T_e = 1$ to 10 eV. These results also confirm the phenomenon  of phonon hardening~\cite{Recoules06}.
 It should be noted that the HNC-SSF value is considerably lower than the value from other methods, illustrating
 the importance of using a bridge term in the integral equation for the ion distribution at these coupling strengths.
\subsection{Quasi-Equation of State}
\label{QEOS}
\begin{figure}[h]
\includegraphics[width=\columnwidth]{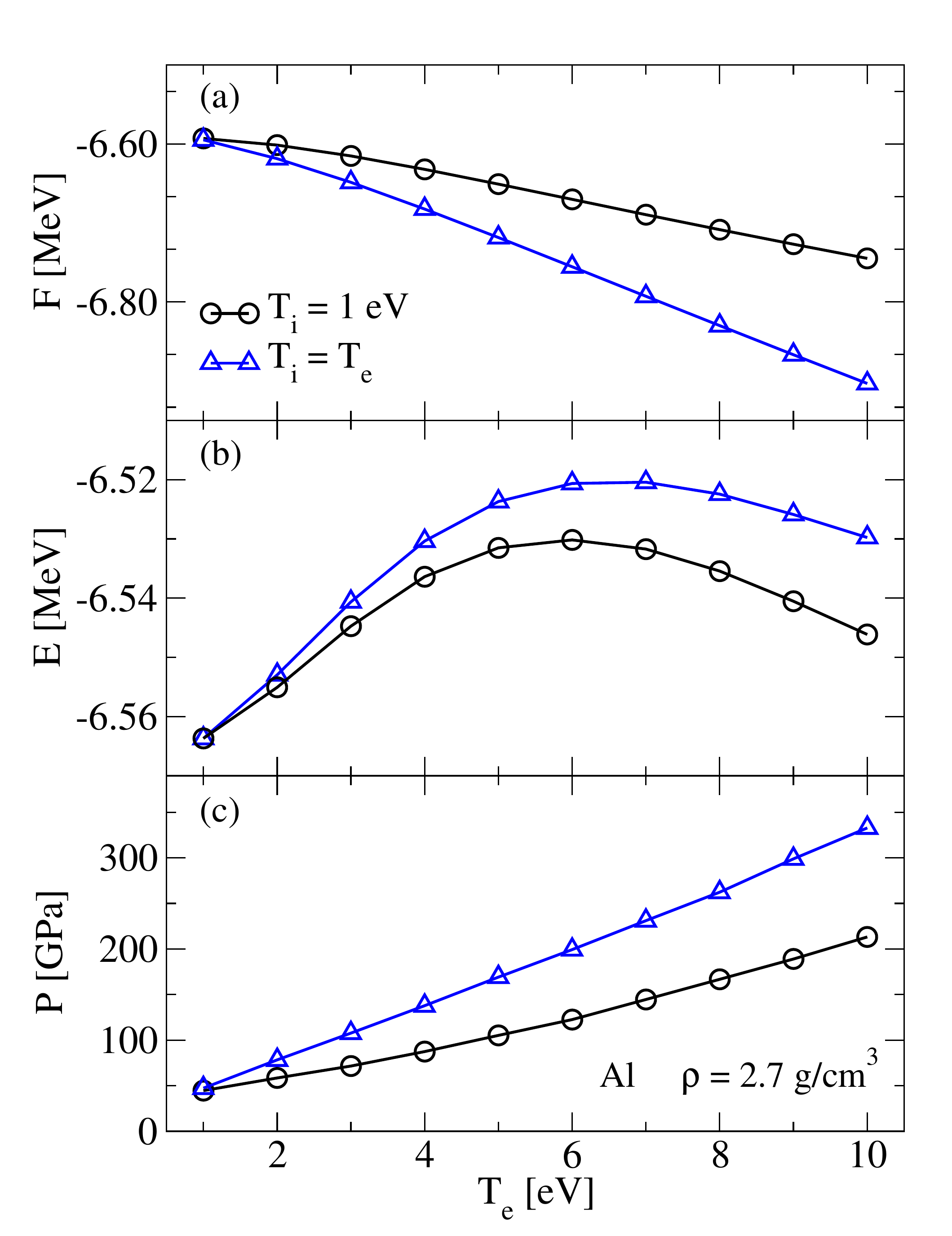}
\caption{(color online) Comparison between the equilibrium and quasi-equilibrium free energy (a),
 internal energy (b), and pressure (c) of Al at  density 
 $\rho_0$. As $T_i$ is held fixed at 1 eV for quasi-equilibrium time scales (or longer via a thermostat
coupled  only to the ions), the ionic lattice does not  expand and the density remains fixed.}
\label{fig:2T.EOS}
\end{figure}

The NPA model allows a rapid calculation of the EOS of Al in equilibrium conditions, which was intensively investigated by Sjostrom \textit{et al} \cite{Sjostrom16}, but also for $2T$ situations. In Fig.~\ref{fig:2T.EOS} we present
 a comparison between  the equilibrium and the quasi-equilibrium Helmholtz free energy $F$, internal energy $E$,
 and total pressure $P$.  At the highest electronic temperature ($T_e = 10$~eV) that we have studied,
 the equilibrium $F$ is lower than that
 of the quasi-equilibrium system by 2.3$\%$  while the internal
 energy is higher by 0.25$\%$. The internal energy in both cases has a maximum in the range of $T_e =6-7$ eV and has
 a similar shape. While $F$ and $E$ are only slightly modified in the $2T$ regime, the equilibrium pressure is higher
 than that obtained at quasi-equilibrium by as much as 56 $\%$ at $T_e$=10~eV. 
Even though the changes in the free energy and internal energy are small, such variations could  considerably
 affect  EOS-dependent properties such as  specific heats, conductivities, energy relaxation
rates   and other coupling coefficients that enter into  more macroscopic WDM  simulations. The efficiency and
rapidity of computing such  $2T$-EOS via the NPA model allows to obtain them on the fly for simulations of
 shocked or  laser-driven systems, for most combinations of $T_i$ and $T_e$ where a significant density of
 free electrons is available to make the NPA approach valid,
 and where no persistent chemical bonds are formed.

\section{Conclusion}
Taking aluminum as an example, we demonstrated that the NPA pair potentials can be used to compute efficiently
and accurately  the equilibrium
 dynamic structure factor via MD simulations, and established that it is in close agreement with DFT-MD results.
 We explored the  two-temperature system and showed that all ion-ion correlations vanish in a time shorter
 than  typical electron-ion relaxation times, validating the concept of a  $2T$-dynamic structure factor
 in this context. 

We presented the $2T$-DSF and showed that the Brillouin peak shifts to higher energies as the electron
 temperature is increased.  As a result, the  ion acoustic mode dispersion relation is modified and the
 adiabatic  speed of sound $c_s$ is  increased, in good agreement with  its determination via the
 compressibility sum rule in the small-$k$ limit  of the static structure factor. The latter is independently
 obtained via the modified hyper-netted chain method and
 using the pair potentials generated via the neutral-pseudoatom method. The increase in the acoustic
 velocity is also consistent with the phenomenon of `phonon hardening'. 

The comparison between the equilibrium and quasi-equilibrium EOS shows that the  free energy and the
 internal energy  are only weakly modified  in the two-temperature system, while the pressure is
 significantly affected. The efficient calculation of the quasi-equilibrium EOS via the neutral
 pseudoatom method  constitutes  a powerful tool for exploring  out-of-equilibrium systems via
 MD simulations.
\section*{Acknowledgments} This work was supported by grants from the
 Natural Sciences and Engineering Research Council of Canada (NSERC)
 and the Fonds de Recherche du Qu\'{e}bec - Nature et Technologies (FRQ-NT). Computations were made on the supercomputer Briar\'{e}e, managed by Calcul Qu\'{e}bec and Compute Canada. The operation of this supercomputer is funded by the Canada Foundation for Innovation (CFI), the minist\`{e}re de l'\'{E}conomie, de la science et de l'innovation du Qu\'{e}bec (MESI) and the Fonds de recherche du Qu\'{e}bec - Nature et technologies (FRQ-NT)

\end{document}